\documentstyle[mncite,draft,epsf]{mn}

\newcommand{\rot}{$\upsilon$~sin$i$ }

\newcommand{\bvec}[1]{\mbox{$\bf{#1}$}}

\newcommand{\ha}{H$\alpha$ }

\newcommand{\caiihk}{Ca\,{\footnotesize II}~H \&~K }

\newcommand{\mgiihk}{Mg\,{\footnotesize II}~h \&~k }

\newcommand{\xbol}{$L_{\rm \small x}/L_{\rm \small bol}$ }

\newcommand{\km}{km~s$^{-1}$ }

\newcommand{\p}{$\pm$ }

\newcommand{\eg}{{\em e.g.,} }

\newcommand{\lsi}{$\mathrel{\hbox{\rlap{\hbox{\lower2pt\hbox{$\sim$}}}\raise2pt
\hbox{$<$}}}$}
\newcommand{\gsi}{$\mathrel{\hbox{\rlap{\hbox{\lower2pt\hbox{$\sim$}}}\raise2pt
\hbox{$>$}}}$}

\title[Super-saturation in M-dwarfs]{X-ray Emission from 
Nearby M-dwarfs: the Super-saturation Phenomenon}

\author[David J. James et al.] 
{David J. James$^{1,4}$\thanks{djj@st-and.ac.uk}, Moira M. 
Jardine$^{1}$, Robin D. Jeffries$^{2}$, Sofia Randich$^{3}$, 
\newauthor Andrew Collier Cameron$^{1}$ \& Miguel Ferreira$^{1,5}$ \\
$^{1}$School of Physics and Astronomy, North Haugh, University 
of St Andrews, St Andrews, Fife, KY16 9SS, UK \\
$^{2}$Department of Physics, Keele University, Keele, Staffordshire, 
ST5 5BG, UK\\ 
$^{3}$ Osservatorio Astrofisico di Arcetri, Largo E.~Fermi 5, 
50125 Firenze, Italy.\\
$^{4}$ Observatoire de Gen\`{e}ve, Chemin des Maillettes 51, CH-1290 
Sauverny, Switzerland.\\
$^{5}$ Departamento de Ci\^{e}ncias Agr\'{a}rias, Universidade dos 
A\c{c}ores, Portugal.}
       
\date{Accepted 2000 December 25.
      Received 2000 December 24;
      in original form 2000 January 01}

\pagerange{\pageref{firstpage}--\pageref{lastpage}}
\def\LaTeX{L\kern-.36em\raise.3ex\hbox{a}\kern-.15em
    T\kern-.1667em\lower.7ex\hbox{E}\kern-.125emX}

\pubyear{2000}

\begin{document}

\label{firstpage}

\maketitle

\begin{abstract}
A rotation rate and X-ray luminosity analysis is presented for rapidly 
rotating single and binary M-dwarf systems. X-ray luminosities for the 
majority of both single \& binary M-dwarf systems with periods below 
$\simeq 5-6$ days (equatorial velocities, V$_{eq}$\gsi 6 km~s$^{-1}$) 
are consistent with the current rotation-activity paradigm, and appear 
to saturate at about $10^{-3}$ of the stellar bolometric luminosity. 

The single M-dwarf data show tentative evidence for the super-saturation 
phenomenon observed in some ultra-fast rotating (\gsi 100 km~s$^{-1}$) 
G \& K-dwarfs in the IC 2391, IC 2602 and Alpha Persei clusters. The IC 
2391 M star VXR60b is the least X-ray active and most rapidly rotating 
of the short period (P$_{rot}$\lsi 2 days) stars considered herein, with 
a period of 0.212 days and an X-ray activity level about 1.5 sigma below 
the mean X-ray emission level for most of the single M-dwarf sample. For 
this star, and possibly one other, we cautiously believe that we have 
identified the first evidence of super-saturation in M-dwarfs. If we are 
wrong, we demonstrate that only M-dwarfs rotating close to their break 
up velocities are likely to exhibit the super-saturation effect at 
X-ray wavelengths. 

The M-dwarf X-ray data also show that there is no evidence for any difference 
in the X-ray behaviour between the single and binary systems, because for 
the single stars, the mean log \xbol $= -3.21 \pm 0.04$ (0.2\lsi P$_{rot}$\lsi 
10.1 days), whereas for the binary stars, the mean log \xbol $= -3.19 \pm 
0.10$ (0.8\lsi P$_{rot}$\lsi 10.4 days).

Furthermore, we show that extremely X-ray active M-dwarfs exhibit a 
{\em blue excess} of about 0.1 magnitudes in U$-$B compared to less 
active field M-dwarfs. Such an excess level is comparable 
to that observed for extremely chromospherically active M-dwarfs. 
Moreover, as is the case for M-dwarf \caiihk activity levels, there 
is an {\em exclusion zone} of X-ray activity between the extremely 
active M-dwarfs and the less active ones.
\end{abstract}


\section{Introduction}

It is now widely believed that radiative losses from the outer 
atmospheres of solar-type stars (spectral types late-F$\rightarrow$M) 
are produced by confinement and heating of stellar plasma in complex 
magnetic field structures above their surfaces (\pcite{schwarz48}; 
\pcite{ulm67}; Rosner et al 1978a, b; \pcite{hey83}). Optical, UV, 
EUV and X-ray observations of solar-type, low mass stars both in the 
field and young (\lsi 700 Myr) open clusters have provided evidence 
for a correlation of increased magnetic activity manifestations 
(\eg H$\alpha$, Ca{\footnotesize II} H \& K, \mgiihk and X-ray fluxes) 
with increasing rotation rate (\eg \pcite{vilhu84}; \pcite{n84}; 
\pcite{doyle87}; \pcite{sod93pl} - S93; \pcite{stauff94}; \pcite{randx-per}). 
The so-called rotation-activity paradigm evident from such observations 
is most-likely a consequence of the stellar {\em dynamo} process which 
converts motions of conducting plasma, in the remnants of a presumably 
relic field, into electric currents and induced magnetic fields. These 
field lines are supposed to be {\em frozen} into the conducting plasma 
(\pcite{ferraro}), and relic field lines threading the stellar convective 
zone are twisted, stretched and deformed by the interaction of differential 
rotation and convective motions leading to the regeneration and enhancement 
of the existing magnetic field. 

The efficiency of the stellar dynamo is related to both the rotation period, 
$P$, and the convective turnover time, $\tau_{c}$, through the Rossby 
number (R$_{o}=P/\tau_{c}$). Magnetic flux, and hence magnetic 
activity induced emissions, increase with decreasing Rossby number (and 
as $\tau_{c}$ is large in lower mass stars, the dynamo is expected to 
be more efficient in the lower mass of two stars having identical 
rotation periods). However there are observational data which suggest, 
at first glance, that the dynamo process does not continue to operate 
{\em ad infinitum}, but {\bf saturates} above some limiting rotational 
velocity (or below a fiducial Rossby number). The data quite clearly 
show that for late-type stars rotating above $\simeq 15-20$ km~s$^{-1}$ 
(more like 5-8 km~s$^{-1}$ for M-dwarfs), a saturation-like plateau in the 
observed chromospheric and coronal emissions is measured (\pcite{vilhu87}; 
S93; \pcite{stauff94}; \pcite{stauff97}). The saturated X-ray 
emission plateau is characterized by X-ray luminosities at about $10^{-3}$ 
of the stellar bolometric luminosities. Moreover, saturation-like emission 
plateaux are also observed in chromospheric and transition region lines, although 
the evidence is less compelling than that for coronal emission (\pcite{vilhu87}; 
S93; \pcite{me-m7}); nonetheless, saturation of the radiative losses 
from relatively rapidly rotating late-type stars appears to be occurring from all 
levels of their outer atmospheres. 

Surprisingly perhaps, the IC 2391, IC 2602 and Alpha Persei open clusters show 
a downward trend from a saturated X-ray emission plateau - termed 
{\bf super-saturation} - for a few G and K-dwarf members with extremely high 
rotation rates (\rot$\sim 100-200$ km~s$^{-1}$, or log R$_{o}$ between $-1.8$ 
and $-2.0$, \pcite{pros96}; \pcite{rancsss10}). Unfortunately there are too few 
stars in these authors' samples exhibiting super-saturation to say whether this 
occurs at a given rotation rate or specific Rossby number. This is especially 
apparent for the Rossby number range as the Randich (1998) Rossby numbers are 
taken from Stauffer et al. (1997), who in turn use rotation period estimates 
using a projected equatorial velocity (\rot) and a mean inclination angle of 
$4/\pi$ . However a {\em lower} limit for equatorial velocities of about 100 \km 
for super-saturation to occur in these stars can be inferred from their \rot 
values. Similarly, this equates to a {\em upper} limit of Rossby number 
for super-saturation of around $-1.40$ (see discussion).

The physical cause of the super-saturation (and indeed saturation !) of stellar 
X-ray emission is far from being fully understood. It is possible that the dynamo 
itself is self-limiting, maybe via a Lorentz back-reaction and shear stresses across 
field boundaries (\pcite{charm92}), or that a saturation of the magnetic heating 
processes which energize coronal plasma is occurring in the most rapidly rotating 
stars. However, O'Dell et al. (1995) discuss arguments that it is not the dynamo 
that is saturating but the star-spot activity (under the assumption that spot 
coverage acts as a tracer for the total surface magnetic flux). They present 
differential photometry data which indicate that star-spot activity (and hence 
surface magnetic flux) in solar-type stars saturates at a rotation at least 6-10 
times at that inferred from chromospheric \& coronal flux data. Such a hypothesis 
is given credence by the theoretical work of Solanki, Motamen \& Keppens (1997), who 
show that rapid rotation leads to the concentration of magnetic flux near the poles 
of rapidly rotating solar-type stars. Such regions will act to mimic the saturation 
of magnetic activity indicators by creating more open-field regions (reducing the 
trapping and heating of upper atmosphere plasma). This effect also provides an 
acceptable alternative for stellar spin-down models which require dynamo saturation 
to fit the rotation-age data for young open clusters.

An alternative hypothesis is that the X-ray emitting coronal volume of rapid 
rotators is reduced via centrifugal stripping (\eg \pcite{moira99}). As the stellar 
rotation rate increases, centrifugal forces cause a rise in the pressure and density 
in the outer parts of the largest magnetic loops, predisposing them to prominence formation 
at the co-rotation radius. This process stresses and distorts the field, opening up 
previously closed field lines (destroying X-ray emitting regions) when the prominences 
erupt. Jardine \& Unruh (1999) showed that if the extent of the corona is limited by the 
co-rotation radius, then X-ray saturation and super-saturation occur naturally for 
G stars,  without the need for dynamo saturation. 

The models of Jardine \& Unruh (1999) show that for increasing rotation rate in G-dwarfs 
(their Fig~1), coronal temperatures increase quite considerably (up to an order of magnitude 
or more). If their models mirror some degree of physical reality, we must then also consider 
the possibility that rapid rotators may sustain sufficiently hotter coronae such that the 
{\sc ROSAT} observatory can no longer measure the peak of their emission distributions in 
its relatively narrow 0.1-2.4 keV passband (see also Randich 1998). While there is no 
evidence that coronal temperatures of late-type stars behave anywhere near such extremes 
as ten-fold increases in coronal temperatures for X-ray active stars compared to low 
activity stars, there is evidence of some temperature difference between the coronae 
of X-ray active and more inactive late-type stars. Gagn\'{e}, Caillault \& Stauffer 
(1995) have provided analyses of X-ray data for G, K \& M-dwarf Pleiads which show that 
single and two temperature plasma model fits to the data both yielded moderately hotter 
coronal temperatures for the more rapidly rotating G \& M stars in their sample. While 
we accept there may be some effect from a likely temperature shift in the coronae of 
rapid rotators, we do not believe such an effect can produce the required shift in the 
peak emission measure to explain the observed super-saturation in the rapid rotators 
(see discussion), and some or other hypotheses must be more fully explored. 

We wish to investigate whether super-saturation occurs at a given period, rotation speed 
or Rossby number and thus gain some insight as to the physical mechanism that causes it. 
The key to such an investigation is to sample the extremes of {\em both} rotation rate 
and convection zone depth. For this reason we present an analysis of a sample of single 
and binary M-dwarfs, with thick convection zones, and contrast their behaviour with 
the rapidly rotating, super-saturated G \& K-dwarfs in the IC 2391 and Alpha Per clusters. 
The paper is structured as follows: in \S~2, the physical data, and their sources, are 
listed for our sample together with the analysis procedures adopted. We present the results 
in \S~3, while a fuller discussion of the data and their implications are deferred to 
\S~4. Finally, we summarize and conclude our findings.

\section{The Sample \& Data Processing}

A sample of rapidly rotating single and binary M-dwarfs has been assembled from a 
variety of literature searches and the SIMBAD database. All single stars were chosen 
such that their photometric rotation periods were known, and a {\sc ROSAT PSPC} X-ray 
observation existed for each. The binary stars were chosen rather more selectively. 
Their rotation periods are inferred from the spectroscopically determined orbital 
motion of the binary under the assumption that each component of the binary system 
is tidally-locked to its companion through a tidal coupling of its orbital and 
rotational motion (for orbital periods of $\sim 10$ days or less - Zahn 1977, 1989). 
Binary M-dwarf systems were therefore chosen having periods of $\sim 10$ days or 
less, and if a {\sc ROSAT } X-ray observation existed for each system.

Physical data for the single and binary M-dwarfs are presented in Tables 1, 2 \& 3 
respectively. All available V, B-V and V-I (Kron or Cousins systems) photometric data 
for the single and binary systems are listed in Table 1. The various sources of the 
data are referenced in column 6. For some systems only V-Ik data are available. These 
data were transformed onto the Cousins system using the V-Ik, V-Ic relationship 
presented in Bessel \& Weis (1987). X-ray data are presented in Tables 2 \& 3 for single 
and binary M-dwarfs respectively. X-ray flux has been divided equally between binary 
components. Apart from the Alpha Per (AP) stars (E$_{V-Ic}$=0.125), the IC 2602 (R) 
stars (E$_{V-Ic}$=0.04) and the IC 2391 (VXR) stars (E$_{V-Ic}$=0.01), negligible 
reddening is assumed for all the remaining single and binary stars. Systematic errors 
in the various X-ray datasets can be reduced by removing the uncertainties in stellar 
distance by using \xbol as the X-ray activity parameter. The calculation of bolometric 
luminosities for M stars is obviously sensitive to the choice of bolometric corrections 
[B.C.]. The usual method for solar-type stars is to use a B-V colour relation. However, 
the B-V colours of M-dwarfs are relatively insensitive to temperature and so one must 
use the V-I colour to determine bolometric corrections. The B.C. for all stars presented 
in Tables 2 \& 3 are calculated using Eqn. 6 from Monet et al. (1992) and the Cousins 
V-I photometric data.  For all binary systems, the bolometric luminosity has been 
calculated assuming the V magnitude is 0.$^{m}$75 dimmer (assumes equal mass components).

\begin{table}
\caption[]{\protect \small For all single and binary field M-stars (ie non-cluster stars) 
listed in Tables 2 \& 3, observed V, B-V, V-Ik and V-Ic photometric data are presented in 
columns $2\rightarrow5$ respectively. Sources of the data are listed in column 6. The 
stars with entries in the V-Ik columns do not have V-Ic data available, and the tabulated 
V-Ic values are determined from the V-Ik, V-Ic relationship presented in Bessel \& Weis 
(1987).} 
\label{data-phot}
\begin{tabular}{lrcccc}
 & & & & & \\
 ~~Star & V~ & B-V & V-Ik & V-Ic & Ref \\
 & & & & & \\
%
%
Gl 182 & 10.05 & 1.40 & - & 1.83 & A \\
Gl 285 & 11.19 & 1.6 & - & 2.95 & A \\
Gl 388 & 9.28 & 1.54 & - & 2.50 & B \\
Gl 411 & 9.47 & 1.51 & - & 2.15 & B \\
Gl 551 & 11.11 & 1.97 & - & 3.67 & A \\
Gl 803 &  8.73 & 1.45 & -  & 2.06 & C \\
Gl 873 & 10.23 & 1.61 & 2.57 & 2.68 & D \\
Gl 875.1 & 11.62 & 1.46 & 2.38 & 2.51 & E \\
RE1816+541 & 11.83 & 1.45 & - & 2.01 & F \\
Gl 890 & 10.84 & 1.42 & - & 1.84 & A \\
%
%
%
HD 16157 & 8.81 & 1.36 & - & 1.86 & G \\
Gl 268 & 11.48 & 1.72 & 2.97 & 3.04 & E \\
FF And & 10.53 & 1.50 & 1.95 & 2.10 & D \\
YY Gem & 9.09 & 1.42 & - & 1.87$^{\alpha}$ & H \\
CM Dra & 12.90 & 1.53 & - & 2.92 & A, I \\
FK Aqr & ~9.09 & 1.46 & - & 2.19 & J \\
RX J0222.4 & 11.10 & 1.44 & - & 1.74 & K \\
+4729 & & & & & \\
Gl 841a & 10.45 & 1.49 & - & 2.38 & L \\
 & & & & & \\
\end{tabular}
\begin{flushleft}
{\small \protect NOTES: 
$\alpha -$ Determined from a V-I (Johnson)~=~2.40. It has been transformed to the Cousins 
system using the relationship, V-Ic = 0.835(V-I)j $- 0.13$  (for 2$<$(V-I)j$<$3 $-$ Bessel 1979). \\
Refs: A: Bessel (1990); B: Celis (1986); C: Cutispoto (1998a) \\
D: Weis (1993); E: Weis (1996); F: Schwartz et al. (1995); \\
G: Cutispoto (1998b); H: Barnes, Evans \& Moffet (1978); \\
I: Eggen (1986); J: Cutispoto \& Leto (1997); \\
K: Chevalier \& Ilovaisky (1997); L: Laing (1989).\\}
\end{flushleft}
\end{table}

\begin{table*}
\caption[]{\protect \small Photometric and X-ray data are tabulated for single M-dwarfs
in the field and three young (\lsi 50 My) open clusters with known rotation periods and 
X-ray detections. Apart from the Alpha Per (AP) stars (E$_{V-Ic}$=0.125), the IC 2602 (R) 
stars (E$_{V-Ic}$=0.04) and the IC 2391 (VXR) stars (E$_{V-Ic}$=0.01) (A$_{v}=2.68\times 
$E$_{V-Ic}$), negligible reddening is assumed. Bolometric corrections (B.C.) are determined 
from the V-Ic data (Monet et al. 1992). \\}
\label{data-singleM}
\begin{tabular}{lrcccrcccccc}
 & & & & & & & & & & \\
 ~~Star & V$_{\small 0}$~ & (V-Ic)$_{\small 0}$ & Spectral & Period & $\tau_{c}$~~~ & Log & 
X-Ray & X-ray ct Rate & HR$^{I}$ & B.C. & Log \\
& & & Type & (Days) & (Days) & R$_{o}$ & Ref~~ & (ct/s) & & & \xbol$^{II}$ \\
 & & & & & & & & & & \\
%
Gl 182 & 10.05 & 1.83 & dM0.5e & 4.56 & 26.18 & -0.76 & RASS & 0.6514 \p 0.0421 & - & -1.31 & -3.30 \\
Gl 285 & 11.19 & 2.95 & dM4.5e & 2.78 & 27.93 & -1.00 & S95 & 1.4172 \p 0.0727 & -0.22 & -2.60 & -2.95 \\
Gl 388 & 9.28 & 2.50 & dM4.5e & 2.70 & 27.39 & -1.01 & S95 & 3.6709 \p 0.0041 & -0.08 & -2.04 & -3.04 \\
Gl 411 & 9.47 & 2.15 & dM2e & 48.0 & 27.13 & ~0.25  & S95 & 0.1823 \p 0.0043 & -0.63 & -1.64 & -4.30 \\
Gl 551 & 11.11 & 3.67 & dM5e  & 42.0 & 31.46 & ~0.13 & S95 & 1.4102 \p 0.0591 & -0.36 & -3.63 & -3.45 \\
Gl 803 & 8.73 &  2.06 & dM0e  & 4.87 & 26.61 & -0.74 & RASS & 5.9520 \p 0.1210 & - & -1.54 & -2.97 \\
Gl 873 & 10.23 & 2.68 & dM4.5e & 4.38 & 28.02 & -0.81 & S95 & 5.2232 \p 0.0956 & -0.16 & -2.26 & -2.61 \\
%
Gl 875.1 & 11.62 & 2.51 & dM3.5e & 1.64 & 26.69 & -1.21 & RASS & 0.4689 \p 0.0313 & - & -2.05 & -3.12 \\
RE1816+541 & 11.83 & 2.01 & dM2e & 0.46 & 26.61 & -1.76 & RASS & 0.2920 \p 0.0130 & - & -1.49 & -3.01 \\
Gl 890 & 10.84 & 1.84 & dM2e & 0.44 & 26.35 & -1.78 & RASS & 0.4266 \p 0.0586 & - & -1.32  & -3.18 \\
%
%
AP 60 & 15.40 & 2.40 & dM3 & 0.318 & 27.93 & -1.94 & R96 & 0.0040 \p 0.0010 & - & -1.92 & -3.10 \\
AP 96 & 14.21 & 1.95 & dM0 & 0.346 & 26.35 & -1.88 & P96 & 0.0060 \p 0.0010 & - & -1.43 & -3.20 \\
AP 211 & 14.73 & 1.78 & dM0 & 0.288 & 26.10 & -1.96 & P96 & 0.0038 \p 0.0005 & - & -1.26 & -3.12 \\
R24A & 14.52 & 1.82 & - & 1.25 & 26.10 & -1.32 & R95 & 0.0090 \p 0.0010 & - & -1.29 & -3.07 \\
R26 & 15.08 & 2.11 & - & 5.70  & 27.04 & -0.68 & R95 & 0.0023 \p 0.0006 & - & -1.59 & -3.56 \\
R31 & 15.00 & 2.20 & - & 0.49  & 27.48 & -1.75 & R95 & 0.0110 \p 0.0010 & - & -1.69 & -2.95 \\
R32 & 15.01 & 2.12 & - & 4.00  & 27.84 & -0.84 & R95 & 0.0063 \p 0.0009 & - & -1.61 & -3.15 \\
R44 & 14.73 & 1.99 & - & 5.50  & 27.13 & -0.69 & R95 & 0.0026 \p 0.0007 & - & -1.47 & -3.59 \\
R50 & 14.63 & 2.03 & - & 6.40  & 27.21 & -0.63 & R95 & 0.0080 \p 0.0010 & - & -1.51 & -3.16 \\
R53B & 15.25 & 2.45 & - & 0.41 & 27.66 & -1.83 & R95 & 0.0079 \p 0.0008 & - & -1.98 & -3.11 \\
R56 & 13.64 & 1.56 & - & 4.10  & 26.10 & -0.80 & R95 & 0.0140 \p 0.0010 & - & -1.05 & -3.13 \\
R57 & 15.43 & 2.40 & - & 8.70  & 27.57 & -0.50 & R95 & 0.0041 \p 0.0007 & - & -1.92 & -3.30 \\
R77 & 14.08 & 1.68 & - & 10.10 & 26.44 & -0.42 & R95 & 0.0031 \p 0.0009 & - & -1.16 & -3.65 \\
VXR38a & 13.34 & 1.57 & dK7.5e & 2.78 & 24.87 & -0.95 & PS96 & 0.0198 \p 0.0012 & - & -1.06 & -3.33 \\
VXR41  & 13.55 & 1.54 & dK7.5e & 5.80 & 25.03 & -0.64 & PS96 & 0.0206 \p 0.0011 & - & -1.03 & -3.22 \\
VXR42a  & 15.86 & 2.48 & dM3e & 1.81 & 27.39 & -1.18 & PS96 & 0.0102 \p 0.0009 & - & -2.01 & -3.00 \\
VXR47  & 13.94 & 2.06 & dM2e & 0.258 & 26.52 & -2.01 & PS96 & 0.0168 \p 0.0015 & - & -1.54 & -3.36 \\
VXR64a  & 15.30 & 1.82 & - & 0.543 & 26.27 & -1.68 & PS96 & 0.0060 \p 0.0004 & - & -1.29 & -3.16 \\
VXR60a  & 14.43 & 2.09 & - & 0.930 & 26.78 & -1.46 & PS93 & 0.0092 \p ----- & -  & -1.57 & -3.44 \\
VXR60b  & 13.82 & 1.70 & - & 0.212 & 25.93 & -2.09 & PS93 & 0.0092 \p ----- & -  & -1.18 & -3.52 \\
 & & & & & & & & & & \\
\end{tabular}
\begin{flushleft}
{\small \protect NOTES: $I - $ HR: Hardness Ratio; $II - $ X-ray fluxes in the $0.1-2.4$ keV 
{\sc ROSAT} bandpass. \\
See text for rotation period references. \\
Sources of X-ray detection are listed in column 8, and are indicated as {\bf RASS}: {\em 
ROSAT} All-Sky Survey; {\bf S95}: Schmitt, Fleming and Giampapa (1995); {\bf R96}: Randich 
et al. (1996); {\bf P96}: Prosser et al. (1996); {\bf R95}: Randich et al. (1995); {\bf PS96}: 
Patten \& Simon (1996); {\bf PS93}: Patten \& Simon (1993). \\
Five separate {\em ROSAT PSPC} flux conversion factors have been used to calculate 
X-ray fluxes. The {\em standard} RASS CF, for \\
negligible interstellar absorption, of $6\times 10^{-12}$ erg cm$^{-2}$ ct$^{-1}$ has been 
used for RASS sources. For Schmitt et al. (1995) sources, a \\ 
CF of ($5.30\times$HR $+8.31$) $\times 10^{-12}$ erg cm$^{-2}$ ct$^{-1}$ has been used, 
whereas the CF used for Alpha Persei (AP) stars is $2\times10^{-11}$ erg cm$^{-2}$ ct$^{-1}$, 
and a CF of $1.2\times10^{-11}$ erg cm$^{-2}$ ct$^{-1}$ is used for the IC 2602 (R) stars. \\
For IC 2391 (VXR) stars, the PSPC CF used was $7.1\times10^{-12}$ erg cm$^{-2}$ ct$^{-1}$. \\
For four of the IC 2391 stars, 42a, 60a,b, 64a, there are no B-V data values available. To 
calculate Rossby numbers for these stars, a field-star V-I to B-V relation was first used 
(Caldwell et al. 1993). }
\end{flushleft}
\end{table*}

Spectral-type data for all single and binary stars are taken from the photometry references 
in Table 1 and/or the SIMBAD database and references therein. For the three Alpha Persei 
stars (AP) only V-I Kron values were available. These data were transformed onto the Cousins 
system using the same V-Ik, V-Ic relationship employed for the field stars. Rotation periods 
for the single stars, column 5 Table 2, are taken from the following:- Gliese star data $-$  
Mathioudakis et al. (1995); Alpha Persei (AP) data $-$ Prosser et al. (1993), Prosser \& Grankin 
(1997); IC 2602 (R) data $-$ Barnes et al. (1999), IC 2391 (VXR) data $-$ Patten \& Simon (1996), 
Simon \& Patten (1998), whereas the RE1816+541 datum is from Robb \& Cardinal (1995). 
Orbital/rotation periods for the binary M-dwarf sample, column 5 Table 3, are taken from 
Jeffries \& Bromage (1993), Jeffries et al. (1993), Chevalier \& Ilovaisky (1997) and Dempsey 
et al. (1997). In columns 6 \& 7 of Tables 2 \& 3 convective turnover times ($\tau_{c}$) and 
Rossby numbers (R$_{o}$) are presented. Convective turnover times (in days) are calculated 
from Eqn.~4 presented in Noyes et al. (1984), suitable for stars with B-V $>$ 1.00, with an 
assumed mixing length parameter of $\alpha=2$. Sources of X-ray detection for the single stars 
are detailed as footnotes to Table~\ref{data-singleM}. Five separate {\em ROSAT PSPC} flux 
conversion factors {[CFs]} have been used to calculate X-ray fluxes for the single stars, 
however only minor differences in absolute X-ray flux levels will result. 

\begin{table*}
\caption[]{\protect \small  Photometric and X-ray data are tabulated for binary M-dwarfs
in the field with known rotation periods (below $\sim$ 10 days) and X-ray detections (in 
the {\em ROSAT PSPC} $0.1-2.4$ keV passband). Negligible reddening is assumed, and bolometric 
corrections (B.C.) are determined from the V-Ic data. \\}
\label{data-binaryM}
\begin{tabular}{llccccccccc}
 & & & & & & & & & \\
~~Star$^{\ast}$ & V$_{\small 0}$ & (V-Ic)$_{\small 0}$ & Spectral & Period & $\tau_{c}$ & 
Log & X-Ray & X-ray ct Rate$^{\alpha}$ & B.C. & Log \\
& & & Type & (Days) & (Days) & R$_{o}$ & Ref & (ct/s) & & \xbol$^{\beta}$ \\
 & & & & & & & & & & \\
%
Gl 841a & 10.45 & 2.38 & dM3-5e + dM3-5e & 1.124 & 26.95 & -1.38 & RASS & 1.0770 \p 0.0812 & -1.90 & -3.16 \\
Gl 268 & 11.48 & 3.04 & dM5e + dM5e & 10.43 & 29.03 & -0.44 & RASS & 0.2570 \p 0.0410 & -2.72 & -3.70 \\
FF And & 10.53 & 2.10 & dM1e + dM1e & ~2.17 & 27.04 & -1.10 & RASS & 1.0220 \p 0.0473 & -1.58 & -3.03 \\
YY Gem & 9.09 & 1.87 & dM1e + dM1e & ~0.81 & 26.35 & -1.51 & RASS & 3.6970 \p 0.0912 & -1.35 & -2.95 \\
CM Dra &  12.90 & 2.92 & dM4e + dM4e & ~1.27 & 27.30 & -1.33 & RASS & 0.1767 \p 0.0155 & -2.56 & -3.23 \\
FK Aqr &  ~9.09 & 2.19 & dM2e + dM3e & ~4.39 & 26.69 & -0.78 & RASS & 3.8030 \p 0.1869 & -1.68 & -3.07 \\
\hline \hline
HD 16157 & 8.81 & 1.86 & dK7 + dM3 & 1.56 & 25.85 & -1.22 & RASS & 3.6790 \p 0.2241 & -1.34 & -3.06 \\
RX J0222.4 & 11.10 & 1.74 & dM0e + dM5e & ~0.465 & 26.52 & -1.76 & RASS & 0.2211 \p 0.0223 & -1.21 & -3.32 \\
+4729 & & & & & & & & & &  \\
 & & & & & & & & & \\
\end{tabular}
\begin{flushleft}
{\small \protect NOTES: All X-ray detections are from the {\em ROSAT} All-Sky Survey and the {\em RASS} 
CF, for negligible interstellar absorption, \\ 
of $6\times 10^{-12}$ erg cm$^{-2}$ ct$^{-1}$ has been used. \\
The two systems detailed below the two horizontal lines are included for reference only, 
and will not be included in the discussion (see text).\\
$\alpha -$ X-ray flux has been divided equally between the components. \\
$\beta -$ Bolometric flux has been been calculated using V$ + 0.^{m}75~-$ assuming 
two equal mass components. \\
$\ast$ - Alternative names: FF And $\equiv$ GJ 9022; YY Gem $\equiv$ Gl 278c; CM Dra 
$\equiv$ Gl 630.1a; FK Aqr $\equiv$ Gl 867a \\}
\end{flushleft}
\end{table*}


The source of X-ray detections for the binary M stars is listed in column 8 of Table 3, 
and all are sources from the {\em ROSAT} All-Sky Survey. The standard RASS CF has been used 
for each, and X-ray flux has been divided equally between the binary components. The plots 
and the discussion do not include the two binary systems RX J0222.4+4729 and HD 16157 because 
they have unequal mass components and the correct allocation of X-ray and bolometric flux 
is problematic. Bolometric flux for the remaining binary systems has been been calculated 
assuming two equal mass components. Gliese 866 and Gliese 570B have not been included in the 
sample due to their long periods of rotation (2.2 yrs and 309 days respectively). DF UMa has 
also been excluded because the system parameters are not fully understood, especially the 
spectral type of the secondary. The recent discovery of the new M-dwarf eclipsing binary 
GJ 2069A has also been excluded because it is believed to be in a quadruple system of 
M-dwarfs, with another M-dwarf binary GJ 2069B (see Delfosse et al. 1999), and the 
allocation of X-ray flux is problematic. Although RE0618+75 is an excellent candidate 
(P$_{rot}=0.539$ days - Jeffries et al. 1993), there are no presently available photometric 
data for this system. 
%

\section{Results} 

\begin{figure*}
\vspace*{10cm}
\includegraphics{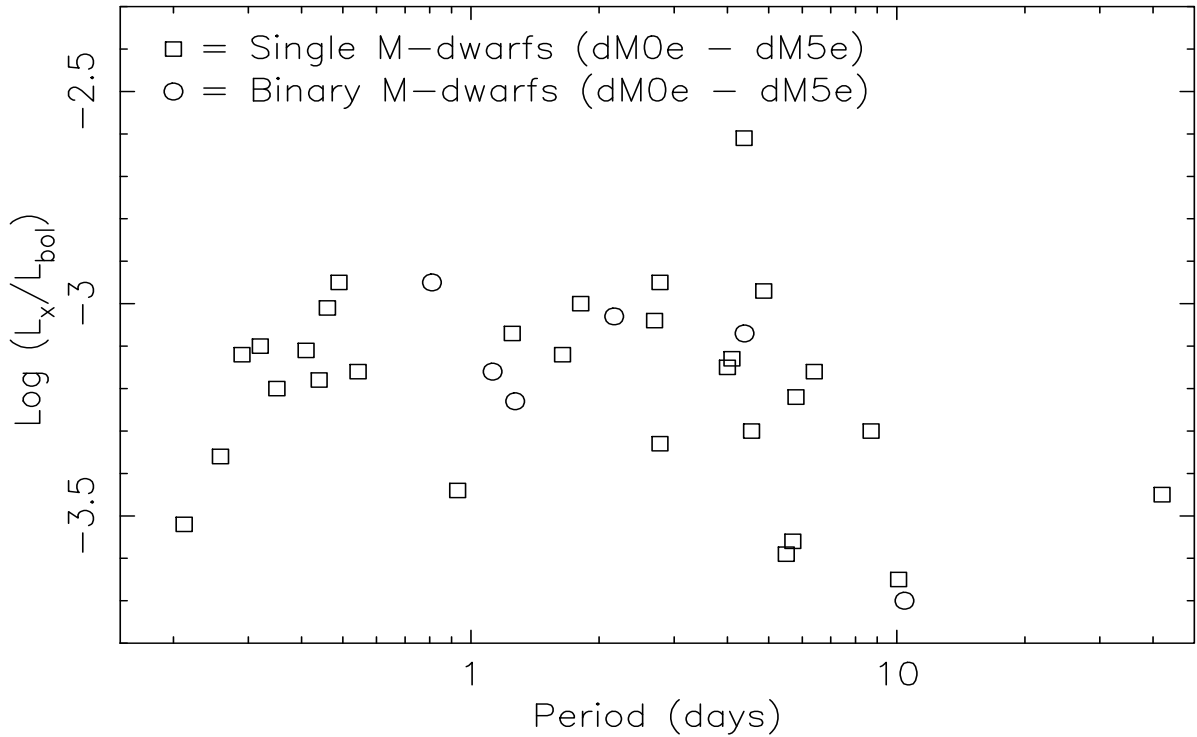}
\caption{\protect \small X-ray luminosity, as a fraction of bolometric 
luminosity, is plotted against rotation period (based on a 24hr day) 
for the single and binary M-dwarfs considered in the present sample (excludes 
RX J0222.4+4729 and HD 16157). The moderately high datum (Gl 873) at log 
\xbol = $-2.61$ is likely due to a flaring event on the star at the time 
of the X-ray observations (see text).}
\label{Mprotplot}
\end{figure*}

\begin{figure*}
\vspace*{10cm}
\includegraphics{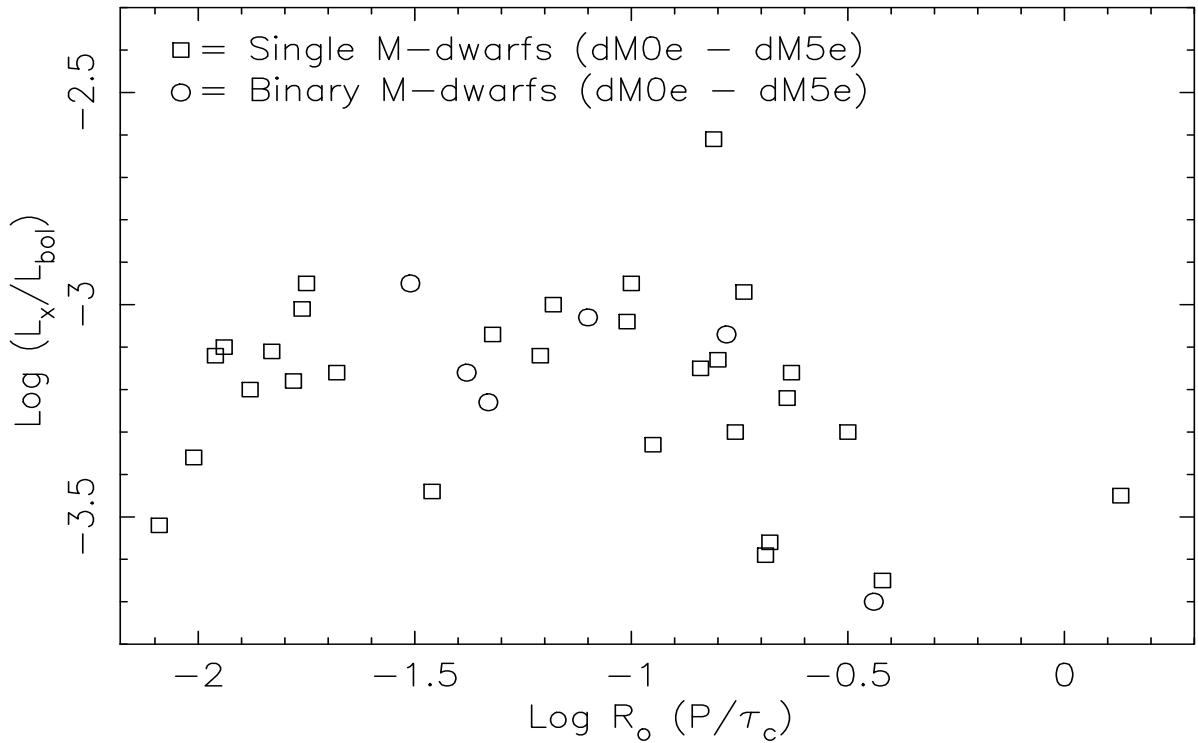}
\caption{\protect \small X-ray luminosity, as a fraction of bolometric 
luminosity, is plotted against Rossby number R$_{o}$ ($\equiv$ 
P$_{rot}$/$\tau_{c}$) for the single and binary M-dwarfs considered in 
the present sample (excludes RX J0222.4+4729 and HD 16157). There is tentative 
evidence for the super-saturation effect at log R$_{o}\sim -2.0$, 
which is also seen in some extremely rapidly rotating IC 2391, IC 2602 and 
Alpha Persei G \& K-dwarfs (\rot$\sim100-200$ \km and log R$_{o}\sim-2.1$; 
Prosser et al. 1996; Randich 1998).}
\label{MRossbyplot}
\end{figure*}

X-ray luminosities, as a fraction of bolometric luminosities, are plotted against 
rotation period and Rossby number for our sample of single and binary M-dwarfs in 
Figures~\ref{Mprotplot} \& ~\ref{MRossbyplot} respectively. We note that the relatively 
high datum at $L_{\rm \small x}/L_{\rm \small bol}=10^{-2.61}$ in both figures is for 
Gl 873, and is moderately higher than the expected X-ray saturated level of $10^{-3}$ 
seen in G, K and M-dwarfs both in the field and young open clusters. The PSPC data 
for Gl 873 have been extracted from the public archive (RP 200984n00 PSPC) and 
a time-series analysis reveals the presence of a steep increase in X-ray photon 
count rate during the exposure, most likely indicating a flaring event in the 
star's corona. In fact, there already exists evidence that this star is highly 
X-ray active and indeed shows flaring activity. X-ray data taken with the {\sc ASCA} 
observatory show that an intense X-ray flare from the star took place on July 
12 1998, which was as luminous as 25$\%$ of its bolometric luminosity and lasted 
over 3 hours (Favata et al. 2000).

Excluding the two slow rotators (periods greater than 40 days) and the highly 
X-ray active star (Gl 873), it is apparent from Figures~\ref{Mprotplot} \& 
\ref{MRossbyplot} that the majority of the remaining single and binary M-dwarfs 
in the present sample, with periods less than $5-6$ days and log Rossby numbers 
less than -0.7, are at or about the X-ray saturation level of \xbol$\sim~10^{-3}$ 
for rapidly rotating late-type stars. There is tentative evidence of a drop in 
X-ray activity among the single stars as one moves to longer period systems (\gsi 6 
days) and higher Rossby numbers (R$_{o}$\gsi -0.7), which can probably 
be attributed to a decline in the magnetic dynamo efficiency through the 
rotation-activity paradigm. The single star and binary system with periods of 
about 10 days, both have log \xbol values which are several times their 
standard deviations below their respective mean values (see below). Unfortunately, 
there are too few data at or about this period to accurately determine the rotation 
rate at which any drop from X-ray saturation occurs.

It is evident from Figures~\ref{Mprotplot}~\&~\ref{MRossbyplot} that the data look 
very similar in period and Rossby number space. This is because the empirical relation 
of Noyes et al. (1984) used to calculate their convective turnover times flattens-off 
at B-V values above 1.2 or so. However the Noyes et al. parameterization gives a 
perfectly acceptable fit out to B-V=1.63 when compared to X-ray \& rotation period 
data for Hyades members and more modern theoretical stellar models (Pizzolato et al. 
1999). Such models yield slightly better fits to the Hyades data, with moderately 
steeper dependences on B-V colour. However, given that many authors still use the 
Noyes et al. relation we retain its use here while reminding readers of its 
potential limitations.


It is also noticeable from Figures 1 \& 2 that for systems with periods \lsi 6 days, 
or R$_{o}$\lsi -0.7, there is no apparent difference between the X-ray behaviour 
of single and binary M-dwarfs (excludes RX J0222.4+4729 and HD 16157). This is also 
apparent statistically, as for single stars (excluding the two systems with periods 
greater than 40 days and the very X-ray active star, GL 873) the mean log 
\xbol is $-3.21 \pm 0.04$ (error on the mean - 0.2\lsi P$_{rot}$\lsi 10.1 days), 
whereas for binary stars (excluding the two unequal mass systems), the mean log \xbol 
is $-3.19 \pm 0.10$ (0.8 \lsi P$_{rot}$\lsi 10.4 days). However, in the absence of 
error bars for the data, the standard deviations [SD] of the samples are likely 
to be better estimates of the error for each datum. For the single stars the SD 
is 0.20, whereas for the binary stars it is 0.25. 
                    
At the ultra fast rotation end of the sample (periods less than 2 day, 
R$_{o}$\lsi $-1.2$), there is considerable scatter in the X-ray level 
of stars in the sample. The use of differing X-ray CFs amounts to 
differences of about 0.1 in log \xbol space and so not all of the scatter 
can be attributed to this effect. Varying levels of stellar photospheric 
spots affecting the V-Ic photometry by differing amounts may have some 
influence, but again this effect should be really rather small in log \xbol 
space. It is quite conceivable that it is caused by X-ray flaring and/or 
activity cycles similar to that observed for our Sun. In fact, the observed 
\xbol scatter is comparable (and perhaps a little less) to that seen in the 
Hyades and Pleiades clusters (\eg Stern et al. 1995; Micela et al. 1999).

At P$_{rot}$\lsi$0.3$ days or R$_{o}$\lsi -2.0, there are two data points which 
show a reduction in X-ray emission compared to the mean, possibly indicating 
X-ray super-saturation. Furthermore, when one considers that the mean log \xbol 
of the single star sample is $-3.21 \pm 0.04$, then the least X-ray active ultra 
rapid rotator, VXR60b, is about 1.5 sigma below the mean X-ray emission level 
for the single star sample as a whole. This star is also the fastest rotator 
in the sample at 0.212 days. 

Any probable case for super-saturation in this M-dwarf sample is based on a 
couple of stars in the ultra rapid rotator domain. Since the crux of the 
argument is most likely to be concerned with the X-ray properties of VXR60b, a 
few words of explanation and discussion are required. Given that both VXR60a and 
VXR60b are located within the same PSPC detect cell, it is prudent to ask how 
appropriate it is to divide equally the X-ray flux between the two stars. Both 
stars are rapidly rotating with periods below one day. However, as VXR60a is 
of lower mass than VXR60b one may hypothesize which star is the most X-ray 
active of the two ? VXR60a will probably have a more active dynamo than VXR60b, 
if indeed they are operating within the same parameters, due to its deeper 
convection zone. But, VXR60b has a greater surface area and therefore it can 
accommodate more magnetic flux, and hence support more X-ray emitting regions 
than VXR60a. Perhaps these two effects cancel each other out to some extent, 
although if both are saturated and the X-ray flux in each would scale as 
the bolometric luminosity, then this would mean that VXR60b would be 1.22 times 
brighter in X-rays than VXR60a.

Even if we were to allocate all of the X-ray flux detected by that PSPC detect 
cell to VXR60b, what value of \xbol would result ? The PSPC count rate of VXR60b 
would be 0.0184 ct/s and using the PSPC CF and the relevant bolometric luminosity, 
an log \xbol $= -3.22$ would result. This value lies at the mean of the single star 
sample as a whole. Of course, it is wholly unreasonable to suggest that VXR60a would 
not emit substantial X-ray emission of its own due to its rapid rotation, and so 
VXR60b must sustain an activity of log \xbol $> -3.22$. Furthermore, if the total 
X-ray flux for the VXR60 detect cell is to be shared between star 60a and 60b in 
the ratios of their bolometric luminosities, then VXR60b would have a PSPC count 
rate of 0.00101 ct/s and VXR60a 0.00083 ct/s. These count rates would result in 
log \xbol values of -3.48 for each star. Therefore, whether we divide the VXR60 
flux equally between stars VXR60a and VXR60b or divide it according to the ratio 
of their bolometric fluxes, VXR60b still exhibits a substantially lower X-ray flux 
than the mean of the single star sample. Moreover, looking at 
Figures~\ref{Mprotplot}~\&~\ref{MRossbyplot} one can see that the X-ray data 
for VXR60a place it in a fairly low position relative to all other data with 
periods less than 2 days. Lowering its X-ray flux even further by sharing the 
X-ray flux of the VXR60 detect cell between VXR60a and b in the bolometric method 
above would only serve to lower its \xbol even further making it appear more 
incongruent with the surrounding data. It is conceivable that it's X-ray emission 
level is indeed higher, moving it more into line with the other data in this region. 
Allocating more of the X-ray flux from the VXR60 PSPC detect cell to VXR60a would 
have the effect of {\em lowering} the X-ray flux for VXR60b further strengthening 
the argument that it exhibits a super-saturated level of X-ray emission.

There are several M-dwarf data in the single star sample taken from IC 2602 and 
IC 2391 X-ray studies which have the label {\em A} or {\em B} associated with their 
listing, indicating that there is more than one optical counterpart in their X-ray 
detect error circles. For all these stars except the already discussed VXR60a,b 
system and the R53 detection, all other associated optical counterparts are photometric 
non-members of their respective clusters. For the R53 detection, it does indeed have 
two photometric cluster members inside its PSPC detect cell. However, R53A has a radial 
velocity about 5 \km away from the cluster mean for IC 2602, and has \ha in absorption 
at a colour where all other members have \ha in emission (Stauffer et al. 1997). Hence, 
this star has doubts cast against its membership of the IC 2602 cluster. We therefore 
consider R53B as the only cluster member inside that PSPC detect cell and hence 
allocate all the X-ray flux in that PSPC detect cell to it.

\section{Discussion}

In order to make a comparison of our data with those of the sample detailed in Randich (1998), 
one must ensure consistent comparison scales. To assist the reader, we provide the relevant 
equatorial velocity (V$_{eq}$) and Rossby numbers for a range of rotation periods for an M2 
dwarf, B-V=1.50, M=0.39M$_{\star}$, R=0.50R$_{\star}$ (Zombeck 1990) in Table~\ref{period-ref}. 
We are now in a position to interchange discussions referring velocities, period and Rossby 
numbers with ease.

\begin{table}
\caption[]{\protect \small Examples of equatorial velocity (V$_{eq}$) and Rossby number 
(R$_{o}$) for a range of rotation periods for an M2 dwarf with mass and radius, 
M=0.39M$_{\star}$, R=0.50R$_{\star}$ (Zombeck 1990).}
\label{period-ref}
\begin{center}
\begin{tabular}{ccc}
 & & \\
P$_{rot}$ & V$_{eq}$ & log R$_{o}$ \\
(days) & \km & (P$_{rot}$/$\tau_{c}$) \\
 & & \\
~10 & 2.53 & -0.43 \\
5.0 & 5.06 & -0.73 \\
1.0 & 25.3 & -1.43 \\
0.5 & 50.6 & -1.73 \\
0.2 & ~126 & -2.13 \\
& & \\
\end{tabular}
\end{center}
\end{table}

Possible clues to the cause of super-saturation would be revealed if one could 
determine the rotation rate or Rossby number at which the decline from X-ray 
saturation set in; i.e, is rotation the key factor or is there a convection zone 
depth dependence as well ? Given that the solar shell-dynamo models requires both 
these parameters to function, if one effect is more dominant that the other, the 
physical mechanism causing super-saturation may become more apparent.

The Alpha Persei, IC2391 and IC2602 sample of G \& K stars in the Randich (1998) sample 
exhibit a decline of X-ray activity from a saturated level at projected equatorial 
velocities of about 100 km~s$^{-1}$ and higher. Given that these measurements contain 
the unknown projection angle $i$, it is not possible to say at what velocity this effect 
becomes apparent. A similar analysis for X-ray activity and Rossby number is presented 
by Stauffer et al. (1997) using these young open cluster data and those of the Pleiades 
and Hyades members as well. Their results show that a decline from X-ray saturation sets 
in at around log R$_{o}\sim -2.0$. However as with the case for \rot, it is still not 
possible to ascertain the Rossby number at which super-saturation sets in for G \& K 
stars because the Stauffer et al. Rossby numbers are calculated from period estimates 
using the projected equatorial velocity and a mean inclination angle of $4/\pi$ . 
However a {\em lower} limit for equatorial velocities of about 100 \km for 
super-saturation to occur in these stars can be inferred from their \rot values. 
Similarly, this equates to a {\em upper} limit of Rossby number for super-saturation 
of around $-1.40$. Therefore, while we cannot infer the exact velocity (or Rossby number) 
at which super-saturation begins for G \& K stars, we can predict a somewhat narrower 
range of velocities of which it is likely to occur.




\subsection{Are M-dwarfs Super-saturated at X-ray Wavelengths?}

In \S~3, we make the claim that VXR60b, and to some extent VXR47, provide tentative 
evidence of super-saturation in M-dwarfs. And, looking at 
Figures~\ref{Mprotplot}~\&~\ref{MRossbyplot}, one can see that the X-ray 
activity-rotation data for VXR60b appear to lie some way below the mean trend 
for the remainder of the data. The obvious counter-arguments are that this datum 
could be erroneous and/or that it is not the only rapidly rotating late-type star 
in its PSPC detect cell making the true X-ray flux allocation hazardous. Without 
further observations of VXR60a,b with an X-ray instrument with improved angular 
resolution it is not possible to correctly determine their relative X-ray levels. 
Note, the HRI on-board ROSAT was not able to spatially resolve these two stars 
(Simon \& Patten 1998). However, with rotation periods below 1 day, it is likely 
that both these stars have extremely active dynamo-induced magnetic activity. The 
position of VXR60a is at the lower edge of the distribution envelope in X-rays 
for its period, however the considerable scatter of X-ray emission in the remainder 
of the period range precludes any necessity to believe that it exhibits unusually 
low X-ray emission levels. 

\begin{figure*}
\vspace*{10cm}
\includegraphics{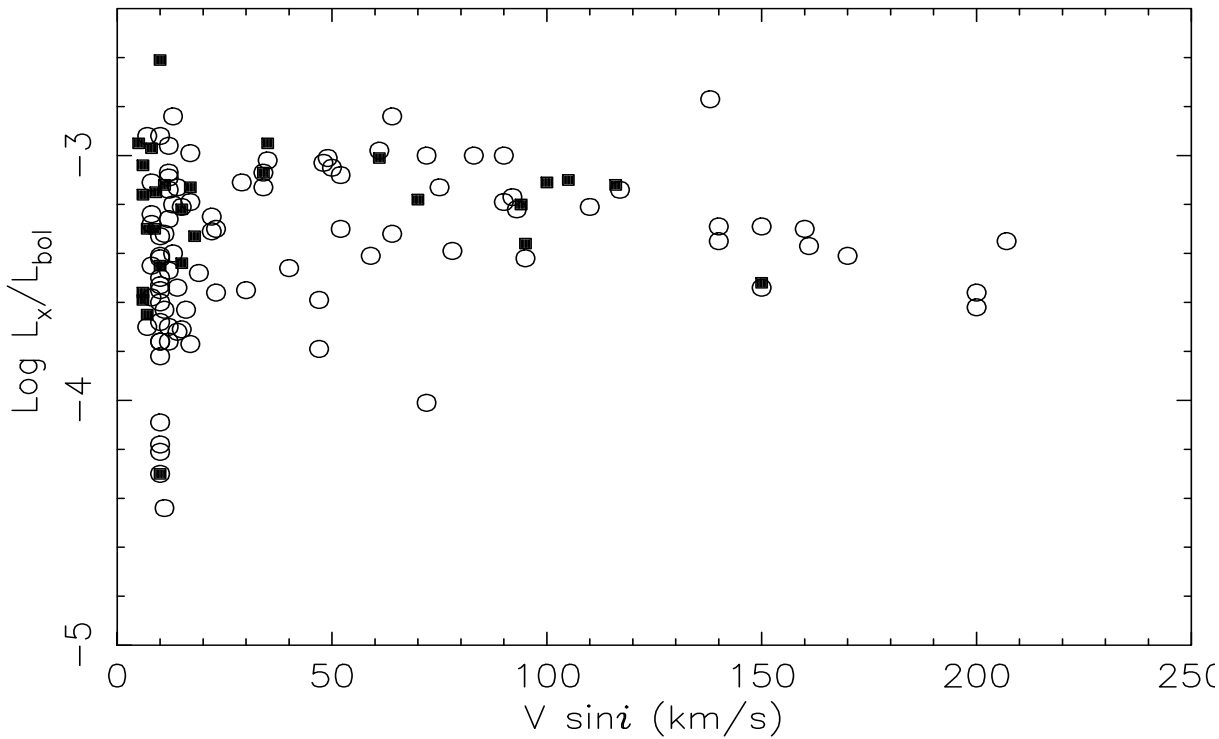}
\caption{\protect \small X-ray luminosity, as a fraction of bolometric 
luminosity, is plotted against projected equatorial velocity (\rot) for 
the single M-dwarfs considered in the present sample (solid squares) 
and for G and K-dwarfs ($0.60<$V-I$_{\small c,o}<1.80$) in the IC 2391/2602 
and Alpha Persei clusters (open circles). The super-saturation effect is the 
trend of decreasing X-ray activity seen among extremely rapidly rotating IC 
2391, IC 2602 and Alpha Persei G \& K-dwarfs (Prosser et al. 1996; Randich 
1998). The M-dwarfs detailed herein appear to the follow a similar trend 
of declining X-ray emission with increasing rotation although there is a 
paucity of M-dwarf data at higher rotation rates. M-dwarf \rot data are taken 
from the same references as for the rotation period or photometric data and/or 
the SIMBAD database.}
\label{vsiniplot}
\end{figure*}

However, if we plot X-ray emission against \rot instead of period for the single M-dwarfs 
in the present sample, further credibility is lent to the existence of the super-saturation 
effect. X-ray luminosity, as a fraction of bolometric luminosity, is plotted against projected 
equatorial velocity for all of the single M-dwarfs considered in the present sample (solid squares) 
and for G and K-dwarfs ($0.60<$V-I$_{\small c,o}<1.80$) in the IC 2391/2602 and Alpha Persei 
clusters (open circles) in Figure~\ref{vsiniplot}. The evidence for X-ray super-saturation 
amongst the G \& K-dwarfs of the three young Galactic open clusters is fairly clear to see 
for \rot \gsi 100 km~s$^{-1}$. Moreover, although the present M-dwarf sample data are less 
abundant in this higher velocity range of the diagram, some semblance of a trend for 
decreasing X-ray emission with increasing rotation rate is visible for the single M-dwarfs. 

If we are to advocate M-dwarf super-saturation then we can say that this effect begins at 
a comparable projected equatorial velocity (100-150 km~s$^{-1}$), and a Rossby number 
(log R$_{o}\sim-2.0$), to the G \& K-dwarfs in the young clusters. On the other hand, if we 
choose to claim that the evidence is too weak to believe that the M-dwarfs in the current 
sample are super-saturated, then it cannot occur at the same rotation rate or Rossby numbers 
as it does for those G \& K stars. Hence, we would be forced to conclude that for rotation 
periods \gsi 0.212 days, M-dwarfs do not exhibit super-saturation at X-ray wavelengths. 
And therefore, any subsequent search for M-dwarf X-ray super-saturation would have to centre 
on stars with periods below $\sim0.21$ days (equatorial velocities greater than $\sim130$ 
km~$^{-1}$). Given that the break-up 
periods\footnote{$\Omega_{br}=(GM_{\star}/4R_{\star}^{3})^{1/2}$} of an 0.39M$_{\star}$, 
0.50R$_{\star}$ M2 dwarf and an 0.47M$_{\star}$, 0.63R$_{\star}$ M0 dwarf are 0.13 \& 0.17 
days, respectively, we would conclude that X-ray super-saturation from M-dwarfs is so 
rare as to be essentially unobservable. 

If we ascertain that M-dwarfs do show super-saturation then an explanation using 
the physical properties of the stars should be put forward. Of course, one must 
be sure that it is indeed an effect of the stellar properties and not the method 
of measurement. In the introduction, we called the readers' attention to the fact 
that G \& M-dwarfs in the Pleiades exhibited single and two temperature plasma 
model fits to {\sc ROSAT} X-ray data both yielding moderately hotter 
coronal temperatures. If one supposes that the super-saturation effect 
is caused by the peak of the X-ray emission moving out of the $1 \pm 1$ 
keV {\sc ROSAT} passband, some semblance of a sensible explanation is 
apparent. However, if one examines the original paper (Gagn\'{e} et al. 
1995) more closely, the case for coronal temperature shifts as an 
explanation for super-saturation is more doubtful. For the composite 
X-ray spectrum of many Pleiads of the same spectral type, the higher 
temperature of their two temperature plasma model fits increase by only 
38 and 21 $\%$  for the G \& M stars, respectively, for those stars with 
\rot  $>$16 km~s$^{-1}$. Furthermore, the absolute levels of the increased 
model temperatures are only kT $= 1.27$ \& 1.15 keV for the rapidly rotating 
G \& M-dwarfs respectively, which is well within the {\sc ROSAT} passband 
(0.1$\rightarrow$2.4 keV). Ultimately, such plasma temperatures are model 
dependent, and our belief that this effect does not contribute significantly 
to the super-saturation phenomenon will only be fully resolved when XMM or AXAF 
observations ($\sim 0.1$$\rightarrow$10 keV passband) for some of these G, 
K \& M-dwarf super-saturated systems are analyzed.

We now explore the hypothesis that the X-ray emitting coronal volume of ultra-rapid 
rotators is reduced via centrifugal stripping, and hence ultra-rapidly rotating 
stars will show decreased levels of X-ray emission, i.e., super-saturation. As the 
stellar rotation rate increases, centrifugal forces cause a rise in the pressure and 
density in the outer parts of the largest magnetic loops. This process can be 
understood in terms of a simple set of arguments. The density structure of a stellar 
corona is determined by a balance between pressure gradients and the {\em effective} 
gravity $\bvec{g_{\rm{eff}}}$ which includes both gravitational and centrifugal terms:

\begin{equation}
 \bvec{\nabla} p =  \rho \bvec{g_{\rm{eff}}}.
\label{hydrostatic}
\end{equation}    

\noindent where $\bvec{\nabla}p$ is the pressure gradient. In the equatorial plane the 
effective gravity has only a radial component ($r$):

\begin{equation}  
\bvec{g_{\rm{eff}}} = \frac{-GM_\star}{ r^2} + \Omega^2 r
\label{geff} 
\end{equation} 

\noindent where $\Omega$ is the rotation rate of the star, and M$_{\star}$is its mass. 
We can integrate (Eqn. \ref{hydrostatic}) analytically for the case where the corona is 
isothermal and composed of an ideal gas with $p = \rho k_B T /m_H$ ($k_B$ is the Boltzmann 
constant and $m_H$ is the hydrogen mass). We then have for a star of radius R$_{\star}$

\begin{displaymath}
\rho(r,\Omega) = \rho_0(\Omega) \: e^{P(r)}
\end{displaymath}
where
\begin{equation}
P(r) = \frac{m_{H}}{kT}
         \left[
             \frac{GM_{\star}}{R_{\star}}
                      \left(
                            \frac{R_{\star}}{r} - 1
                      \right)
           + \frac{\Omega_{\star}^{2} R_{\star}^{2}}{2}
                      \left(
                             \frac{r^{2}}{R_{\star}^{2}} - 1
                      \right)
         \right]
\label{density-full}       
\end{equation}     

The density (and pressure) fall off exponentially with height close to the 
surface, then start to rise again at the co-rotation radius, 
$r_{c} = (GM_{\star}/\Omega^{2})^{1/3}$, where $\bvec{g_{\rm{eff}}}=0$. Eventually, 
the gas pressure becomes greater than the magnetic pressure and the magnetic 
field lines may be blown open.  This height may be taken as the maximum extent 
of the corona.  As the rotation rate increases, the co-rotation radius moves 
closer to the star until it moves inside the maximum extent of the corona.  For 
single G stars, this happens at approximately the same rotation rate at which 
the X-ray emission saturates.  Jardine \& Unruh (1999) show that if the true 
extent of the corona is limited by the co-rotation radius, then the X-ray emission 
saturates naturally without the need for dynamo saturation.  In essence, this is 
because the increase in the emission due to the increase in the density with rotation 
rate is balanced by the decrease in the emitting volume.  If the density varies as 
$n_e\propto p/T$ where $T\propto \Omega$ (Jordan \& Montesinos 1991) 
and from equipartition $p\propto B^2 \propto \Omega^2$ for a simple dynamo 
prescription, then $n_e \propto \Omega$.  If the emitting volume is determined by 
the co-rotation radius, then $V\propto r_{c}^3 \propto \Omega^{-2}$ and so the 
emission measure $\int n_e^2 dV$ becomes independent of the rotation rate.  
This argument breaks down once the co-rotation radius is very close to the 
surface and it is at this point that Jardine \& Unruh found super-saturation.

For M stars, however, with their lower mass and radius, the maximum extent of 
the corona and the position of the co-rotation radius will be different.  How 
does this affect the interpretation of the results presented here?  To illustrate 
this, we take the case of an M2 star of mass $M=0.39 M_\odot$ and radius 
$R=0.50R_\odot$ (Zombeck, 1990) and an M0 star of mass $M=0.47 M_\odot$ and 
radius $R=0.63R_\odot$ (Zombeck, 1990) and compare them to the case of a G star 
of solar mass and radius. At a given rotation rate, the density and pressure 
scale heights depend primarily on the first term in the expression for $P(r)$ 
(Eqn.~\ref{density-full}) and hence on the ratio  $M_{\star}/R_{\star}$.  
For the M2 star, this is only $0.78$ times that for the G star (0.75 times, 
for the M0 star). As a result the pressure scale height will be higher and 
so if the two stars have the same magnetic field, the corona of the M star 
will be smaller. At the same time, the co-rotation radius of the M star will 
be further from the surface since

\begin{displaymath}
r_c/R_\star \propto \frac{M^{1/3}_{\star}}{R_{\star}} \frac{1}{\Omega^{2/3}} \propto 
\frac{(M_{\star}R_{\star})^{1/3}}{V_{eq}}
\end{displaymath}

\noindent
implying that equatorial velocity is the more important parameter in this model. 
Hence, at the same period, the position of the co-rotation radius 
($r_c/R_\star$) for the M2 star will be $1.46$ times that for the G star (1.23 
times, for the M0 star).  As a result, we expect that the intrusion of the 
co-rotation radius into the closed corona, and hence the onset of super-saturation 
of the X-ray emission, will occur at shorter periods in M stars than in G stars. 

In fact, the position of the co-rotation radius for the M and G stars will 
be the same only where the period for the M2 star is 
$(M_{\star}/R_{\star}^{3})^{1/2} = 1.77$ times that for the G star (1.37 times, 
for the M0 star). Among the G \& K-dwarfs of the Alpha Persei cluster, RS 
CVn and W UMa systems (Randich 1998 - her Figure 4) super-saturation becomes 
apparent at log P$_{rot}\sim 0.9$, ie P$_{rot}\sim7.94$ hrs, at which point 
the co-rotation radius is about $1 R_{\star}$ above the stellar surface.            


In order for the M2-dwarf used in the example above to have a rotation rate 
where the co-rotation radius is around $1 R_{\star}$ above the surface, the 
star would need to rotate at periods of around 4.49 hrs (0.19 days) and 
around 5.80 hours, 0.24 days for the M0 star. With a V-Ic,o of 1.70, it is 
likely that VXR60b is nearer to being an M0 dwarf than an M2 dwarf, and so 
referring to Table~2 and Figure~1, one can see that VXR60b has a shorter 
period than 0.24 days, and exhibit a lower level of X-ray emission compared 
to the mean emission levels of the remaining single star sample as a whole 
(excluding the two very long period systems and Gl 873). In fact the least 
X-ray active ultra rapid rotator, VXR60b, is about 1.5 sigma below the mean 
X-ray emission level of the majority of the single star sample.


Unfortunately, we are not able to state that the super-saturation effect occurs 
at a given Rossby number for G, K and M-dwarfs, and so we still cannot rule out 
that this is a property of either the dynamo or the stars themselves. Essentially, 
this is because as the convective turnover time increases, the radius goes down 
by a similar factor, so that for a given equatorial velocity for either, 
the Rossby numbers are roughly the same.

An alternative hypothesis for the explanation of magnetic-activity indicator 
saturation has been suggested, which results in a reduction or absence for the 
need of magnetic dynamo saturation. This hypothesis is based on the premise 
that a larger fraction of the energy budget of the radiative losses from rapidly 
rotating late-type stars is converted into continuum emissions, resulting in a 
{\em blue excess} emission in the colours of such stars (e.g., \pcite{M-models5}; 
\pcite{M-models6}). Such an effect for the chromospheric activity lines \ha and 
Ca\, {\footnotesize II} is predicted by the Houdebine et al. models, and is 
clearly observed among active field M-dwarfs (for a particularly vivid 
demonstration of this effect, see fig. 14c of Houdebine \& Stempels 1997). 

\begin{figure}
\vspace*{7.0cm}
\includegraphics{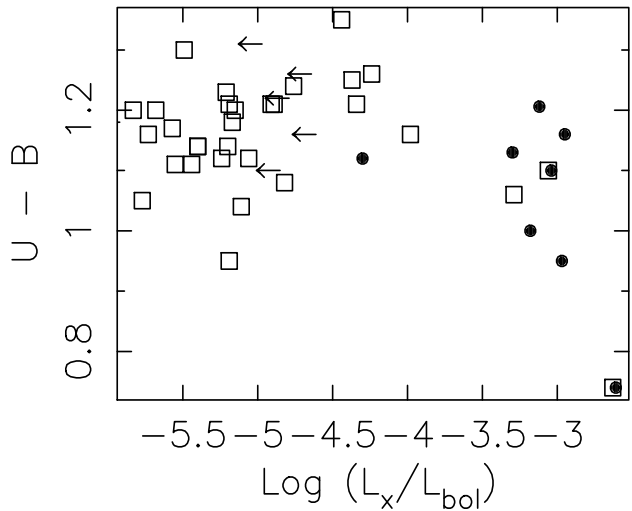}
\vspace*{7mm}
\caption{The photometric colour U-B is plotted against X-ray luminosity, as 
a fraction of bolometric luminosity, for nearby (within 7 parsecs) M-dwarfs 
detected in the RASS (data taken from Fleming et al. 1995 - open squares) 
and the single M-dwarfs (with available colour - filled circles) detailed 
in Table~\ref{data-singleM}. There is statistical evidence of a {\em blue 
excess} (smaller U$-$B values) in the M-dwarfs with saturated levels of 
X-ray emission compared to the less-active, nearby field M-dwarfs. It 
is also clear that there is an {\em exclusion zone} of X-ray activity, 
similar to that noted for chromospheric activity of M-dwarfs in the 
\caiihk lines, between log \xbol of -3.3 and -4.}
\label{UB-excess}
\end{figure}

Recently, J.~Doyle (private communication) has suggested that a similar effect 
may be present in the X-ray domain, implying that the more X-ray active M-dwarfs 
should exhibit a {\em blue excess} in comparison to less active stars of similar 
mass. To this end, we plot X-ray luminosity, as a fraction of bolometric luminosity, 
against U$-$B colour in Fig.~\ref{UB-excess} for a sample of nearby field 
M-dwarfs (spectral type M0V$-$M4V - open squares) and the single M-dwarfs (with 
available colours - filled circles) listed in Table~\ref{data-singleM}. The 
general field star X-ray data are taken from a volume-limited sample (within 
7 parsecs) of M-dwarfs detected in the RASS (Fleming, Schmitt \& Giampapa 
1995), and the U-B colours of these stars and the X-ray active stars listed 
in Table~\ref{data-singleM} are taken from the {\sc SIMBAD} database. 
Unfortunately the open cluster stars listed in Table~\ref{data-singleM} do 
not have any U$-$B data available.

For the RASS detected stars, the mean colour is $<$U$-$B$>~=~1.16 \pm 
0.02$ (error on mean), whereas for the single M-dwarfs exhibiting 
saturated levels of X-ray emission listed in Table~\ref{data-singleM} 
the mean colour is $<$U$-$B$>=1.04 \pm 0.06$ (error on mean, excludes 
two stars with rotation periods over 40 days). Therefore at X-ray 
wavelengths, it appears that there is a {\em blue excess} of about 
0.1 magnitudes in U$-$B for very X-ray active M-dwarfs compared to 
a less-active field star sample. This is a very similar level of U$-$B 
{\em excess} found for extremely chromospherically active M-dwarfs 
(\pcite{M-models6}). Another interesting point of note evident from 
Fig.~\ref{UB-excess} is that there is an {\em exclusion zone} of 
X-ray activity between log \xbol of -3.3 and -4. This could be due 
to a lack of data points in the very X-ray active M-star sample 
detailed in this manuscript. However, a remarkably similar zone is 
observed for chromospheric activity of M-dwarfs in the \caiihk lines. 
If such behaviour at X-ray wavelengths can be successfully modeled, 
the requirement of invoking dynamo saturation to explain magnetic 
activity saturation may be eliminated. 


%

\section*{Summary}

We have presented a rotation rate and X-ray luminosity analyses for rapidly 
rotating single and binary M-dwarf systems. All rotation period and X-ray 
data have been derived from the published literature and the SIMBAD database. 
The X-ray luminosities for the majority of both single \& binary M-dwarf 
systems with periods below $\simeq 5-6$ days (\gsi 6 km~s$^{-1}$) are consistent 
with the current rotation-activity paradigm, and appear to saturate at about 
$10^{-3}$ of the stellar bolometric luminosity. For the ultra-rapid rotators 
with rotation periods \lsi 2 days, there is considerable scatter in the X-ray 
levels observed. This can probably be attributed to magnetic activity 
cycles, flaring or stellar spots. 

The single M-dwarf data show tentative evidence for the super-saturation phenomenon 
observed in some ultra-fast rotating ($100-150$ km~s$^{-1}$) G \& K-dwarfs in 
the IC 2391, IC 2602 and Alpha Persei clusters. This supposition is primarily based 
on one datum for the IC2391 star VXR60b (and possibly VXR 47 as well), a \rot-X-ray 
activity diagram and a theoretical framework which shows a reduction in the coronal 
volume by coronal stripping should occur at the rotation period of VXR60b. The M-dwarf 
X-ray data also show that there is no evidence for any difference in the X-ray 
behaviour between the single and binary systems, because for single stars, the mean 
log \xbol is $-3.21 \pm 0.04$, whereas for the binary stars, the mean log \xbol is 
$-3.19 \pm 0.10$. 

The hypothesis that the X-ray emitting coronal volume of rapid rotators is 
reduced via centrifugal stripping has been explored. We demonstrate that an 
M-dwarf sample would need to incorporate targets with periods of around 
P$_{rot}=5.80$ hours (0.242 day) or lower (where the co-rotation radius is 
around $1 R_{\star}$ above the surface) for X-ray super-saturation to be observed. 
The IC 2391 star VXR60b is the least X-ray active and most rapidly rotating of 
the short period (P$_{rot}$\lsi 2 days) systems considered herein, with a period 
of 0.212 days and exhibits an X-ray activity level about 1.5 sigma below the mean 
X-ray emission level for most of the single M-dwarf sample. We cautiously believe 
that we have identified the first evidence of super-saturation in M-dwarfs. If 
we have not, then only M-dwarfs rotating close to their break up velocities are 
likely to exhibit the super-saturation effect at X-ray wavelengths. 

An analysis of X-ray activity and the U$-$B colours of M-dwarfs shows that 
there is a {\em blue excess} of about 0.1 magnitudes in U$-$B for extremely 
X-ray active stars compared to less active ones. Such an excess level is 
comparable to that observed for extremely chromospherically active M-dwarfs. 
This analysis also shows that there is an {\em exclusion zone} of X-ray 
activity between the extremely active M-dwarfs and the less active ones, 
extending over three quarters of a decade in \xbol space.

\section*{Acknowledgments}

The provision of advice and electronic data by John Caldwell 
at the South African Astronomical Observatory and Mike Bessel at Mount 
Stromlo and Siding Spring Observatory is gratefully acknowledged. This 
research has made use of the {\sc SIMBAD} database, operated at the 
Centre de Donn\'{e}es astronomiques de Strasbourg, La France, and the 
Leicester Database and Archive Service at the Department of Physics and 
Astronomy, Leicester University, UK. DJJ also thanks PPARC for a 
post-doctoral research fellowship, and the Royal Society for a European 
research grant. We wish to thank the referee, Dr Gerry Doyle, for his 
careful reading of the manuscript and the important suggestion of 
checking the U$-$B colours. DJJ would like to thank Mar\'{\i}a Pilar 
Escribano Benito for much love and support during this work, and the 
continued positive influences of Mrs J Pryer. The interesting and varied 
discussion sessions with Matt Constable, Iain Colwell, Johnnie Jack and 
Steve Strain were helpful and are gratefully acknowledged.

\nocite{barnes2602-P}
\nocite{barnes78}
\nocite{bessweis87}
\nocite{bess79}
\nocite{bess90}
\nocite{john-saao}
\nocite{celis86}
\nocite{charm92}
\nocite{chevvie}
\nocite{cuti-6}
\nocite{cuti-7}
\nocite{cuti-8}
\nocite{delf99}
\nocite{demp97}
\nocite{doyle87}
\nocite{eggen86}
\nocite{favata2000}
\nocite{ferraro}
\nocite{flem-RASS95}
\nocite{g95}
\nocite{hey83}
\nocite{M-models5} 
\nocite{M-models6}
\nocite{me-m7}
\nocite{moira99}
\nocite{jeff2-93}
\nocite{jeff3-93}
\nocite{JMont91}
\nocite{laing89}
\nocite{mathio95}
\nocite{micela-hripl99}
\nocite{monetBC}
\nocite{mullan84}
\nocite{n84}
\nocite{odell95}
\nocite{PS93}
\nocite{PS96}
\nocite{pizzo-tauc}
\nocite{prosser93}
\nocite{pros96}
\nocite{pros97}
\nocite{randx-2602}
\nocite{randx-per}
\nocite{rancsss10}
\nocite{robb95}
\nocite{rosner78a}
\nocite{rosner78b}
\nocite{sch95}
\nocite{RE1816+54}
\nocite{schwarz48}
\nocite{SP98}
\nocite{sod93pl}
\nocite{solanki97}
\nocite{stauff94} 
\nocite{stauff97}
\nocite{stern-hy95}
\nocite{ulm67}
\nocite{vilhu84} 
\nocite{vilhu87}
\nocite{weis93}
\nocite{weis96}
\nocite{zahn0}
\nocite{zahn1}
\nocite{zom90}

\bibliographystyle{mn}
\bibliography{mnras_journals}

\label{lastpage}

\end{document}